\documentclass[aps,preprint,superscriptaddress,]{revtex4}
\usepackage{graphics}
\usepackage{graphicx}
\begin{document}
\title{Dislocation constriction and cross-slip in Al and Ag: an {\it ab initio} study} 
\author{Gang Lu}
\affiliation
{Department of Physics and Division of Engineering and Applied Science,
 Harvard University, Cambridge, MA 02138}
\author{Vasily V. Bulatov}
\affiliation{Lawrence Livermore National Laboratory, Livermore, CA 94550}
\author{Nicholas Kioussis}
\affiliation{Department of Physics, California State University Northridge,
Northridge, CA 91330}
\begin{abstract} 
\vskip 0.5cm
A novel model based on the Peierls framework of dislocations is 
developed. The new theory can deal with a dislocation spreading at more than 
one slip planes. As an example, we study dislocation cross-slip and 
constriction process of two fcc metals, Al and Ag. The energetic 
parameters entering the model are determined from {\it ab initio} 
calculations. We find that the screw dislocation in Al can cross-slip 
spontaneously in contrast with that in Ag, which 
splits into partials and cannot cross-slip without first being constricted.
The dislocation response to an external stress is examined in detail.
We determine dislocation constriction energy and critical stress for 
cross-slip, and from the latter, we estimate the cross-slip energy barrier
for the straight screw dislocations. 
\end{abstract}
\maketitle
The cross-slip process by which a screw dislocation moves from one 
slip plane to another, plays an important role for plastic deformation
in materials. For example, cross-slip is responsible for 
the onset of stage III of the stress-strain work-hardening curve; it 
is also responsible for the anomalous high temperature yield stress 
peak observed in L$1_2$ 
intermetallic alloys. However cross-slip has been a tough problem
to tackle because it contains both long-ranged elastic interactions 
between dislocation
segments and short-ranged atomic interactions due to the 
constriction process, in which
the two partial dislocations have to be recombined into a screw dislocation
before cross-slip takes place. 

There are currently two theoretical
approaches to study cross-slip. One is based on
the line tension approximation which completely ignores atomic 
interactions \cite{Friedel,Escaig}. 
The other approach is direct 
atomistic simulations employing empirical potentials \cite{Rasmussen,Rao}.
Although the second approach is quite powerful in determining cross-slip transition
path and estimating the corresponding activation energy barrier, it is
time-consuming and critically depends on the accuracy and availability
of the empirical potentials employed in the simulations. In this 
Letter, we present an alternative approach to study cross-slip process 
based on the Peierls framework with {\it ab initio} 
calculations of relevant energetics.  
In fact, there has been a 
resurgence of interest recently in 
applying the simple and tractable Peierls-Nabarro (P-N) model to study 
dislocation core structure 
and mobility in conjunction with {\it ab initio} $\gamma$-surface calculations
 \cite{Joos,Juan,Bulatov,Hartford,Lu1,Lu3}. This approach represents a
combination of atomistic ({\it ab initio}) treatment of interactions 
across the slip plane and
elastic treatment for the continua that are away from the slip plane. 
Therefore this approach is particularly
useful for studying interactions of impurities and dislocations, when
empirical potentials are either not available or not reliable to deal with such
multi-elements systems. However, to date all the models based on the Peierls
framework are only applicable to single slip plane while cross-slip process
requires at least two active intersecting slip planes, i.e., the primary and 
cross-slip planes.
It is the purpose of this Letter to introduce a novel P-N model that involves 
two intersecting slip planes. This development represents the first effort
to extend the P-N model, one of the central themes of dislocation
theory, to more than one slip planes, which opens doors to many
exciting applications. As an example, we shall apply this new model 
to study dislocation constriction and cross-slip process in Al and Ag. 
Not only can this model be used to study dislocation cross-slip,
it can also be applied to examine dislocation junctions and other 
processes involved multiple slip planes. 

We begin by developing an appropriate energy functional for 
a Peierls dislocation at two intersecting slip planes. 
To facilitate presentation, we adopt the following conventions
: In Fig. 1, a screw dislocation placed at the intersection of the 
primary (plane I) and cross-slip plane (plane II) is allowed to spread into the
two planes simultaneously. The $X$ ($X'$) axis represents the glide
direction of the dislocation at the plane I (II). For an fcc lattice, 
the two slip planes are (111) and ($\bar{1}11$), forming an
angle $\theta \approx$ 71$^\circ$. 
The dislocation line is along the [10$\bar{1}$] ($Z$ axis) direction and $L$ 
represents the outer radius of the dislocation beyond which the elastic
energy is ignored. In the spirit of P-N model, the dislocation is represented
as a continuous distribution of infinitesimal dislocations with densities of
$\rho^{\rm I}(x)$ and $\rho^{\rm II}(x')$ on the primary and 
cross slip planes
respectively. Here $x$ and $x'$ are the coordinates of the atomic rows at
the two planes. Following the semidiscrete Peierls framework
developed earlier \cite{Bulatov,Lu1}, we can write the total energy
of the dislocation as 
$U_{tot}=U_{\rm I}+U_{\rm II}+\tilde{U}$. Here $U_{\rm I}$ and $U_{\rm II}$
are the energies associated with the dislocation spread on the plane I and
II, respectively and $\tilde{U}$ represents the elastic interaction energy
between the dislocation densities on planes I and II. $U_{\rm I}$ and 
$U_{\rm II}$ are essentially the same expression given earlier for the single
plane case \cite{Bulatov, Lu1}, while $\tilde{U}$ is a new term and
can be derived from Nabarro's equation for general parallel 
dislocations \cite{Nabarro},    
\begin{eqnarray*}
U_{\rm I(II)}&=&\sum\limits_{i,j}\frac{1}{2}\chi_{ij}\{K_e[\rho^{\rm I(II)}_1(i)
\rho^{\rm I(II)}_1(j)+\rho^{\rm I(II)}_2(i)
\rho^{\rm I(II)}_2(j)]+
K_s\rho^{\rm I(II)}_3(i)\rho^{\rm I(II)}_3(j)\}\\
& & +\sum\limits_{i} \Delta x\gamma_3\left(f^{\rm I(II)}_1(i),f^{\rm I(II)}_2(i),
f^{\rm I(II)}_3(i)\right)
-\sum\limits_{i,l}\frac{x(i)^2-x(i-1)^2}{2}
\rho^{\rm I(II)}_l(i)\tau^{\rm I(II)}_l+Kb^2{\rm ln}L,\\
\tilde{U}&=&-\sum\limits_{i,j}K_s\rho^{\rm I}_3(i)\rho^{\rm p}_3(j)
A_{ij}-\sum\limits_{i,j}
K_e[\rho^{\rm I}_1(i)\rho^{\rm p}_1(j)+\rho^{\rm I}_2(i)\rho^{\rm p}_2(j)]A_{ij}\\
& &-\sum\limits_{i,j}
K_e[\rho^{\rm I}_2(i)\rho^{\rm p}_2(j)B_{ij}+\rho^{\rm I}_1(i)
\rho^{\rm p}_1(j)C_{ij}-
\rho^{\rm I}_2(i)\rho^{\rm p}_1(j)D_{ij}-\rho^{\rm I}_1(i)
\rho^{\rm p}_2(j)D_{ij}]~.
\end{eqnarray*}
Here $f^{\rm I(II)}_1(i)$, $f^{\rm I(II)}_2(i)$ and 
$f^{\rm I(II)}_3(i)$ represent the edge, vertical and screw component
of the general dislocation displacement at the $i$-th nodal 
point in the plane I(II), respectively, while the corresponding 
component of dislocation density in plane I(II) is defined as 
$\rho^{\rm I(II)}(i) = \left(f^{\rm I(II)}(i)-
f^{\rm I(II)}(i-1)\right)/\left(x(i)-x(i-1)\right)$. The  
projected dislocation density $\rho^{\rm p}$(i) is the projection  
of density $\rho^{\rm II}$(i) from plane II onto plane I in order to deal 
with non-parallel components
of displacement. The  
$\gamma$-surface, $\gamma_3$, which in general includes shear-tension coupling
can be determined from {\it ab initio} calculations. $\tau^{\rm I(II)}_l$
is the external stress components interacting with corresponding
$\rho^{\rm I(II)}_l(i)$ ($l$ = 1,2,3), which contributes to the total energy as elastic
work done by the stress \cite{Bulatov}. Dislocation response
to the external stress is achieved by optimization
of $\rho^{\rm I(II)}_l(i)$ at a given value of $\tau^{\rm I(II)}_l$, but 
dislocation core energy as
an internal energy does not include the contribution from the external work.  
$K_e$ and $K_s$ are
the edge and screw components of the general prelogarithmic elastic energy
factor $K$. $\chi_{ij}$, $A_{ij}$, $B_{ij}$, $C_{ij}$ and $D_{ij}$ are
double-integral kernels defined as following:
\begin{eqnarray*}
\chi_{ij}&=&\int\limits_{x_{j-1}}^{x_j}\int\limits_{x_{i-1}}^{x_i}
{\rm ln}|x-x'|dxdx',\\
A_{ij}&=&\int\limits_{x'_{j-1}}^{x'_j}\int\limits_{x_{i-1}}^{x_i}
\frac{1}{2}{\rm ln}(x_0^2+y_0^2)dxdx',\\
B_{ij}&=&\int\limits_{x'_{j-1}}^{x'_j}\int\limits_{x_{i-1}}^{x_i}
{\rm ln}\frac{x_0^2}{x_0^2+y_0^2}dxdx',\\
C_{ij}&=&\int\limits_{x'_{j-1}}^{x'_j}\int\limits_{x_{i-1}}^{x_i}
{\rm ln}\frac{y_0^2}{x_0^2+y_0^2}dxdx',\\
D_{ij}&=&\int\limits_{x'_{j-1}}^{x'_j}\int\limits_{x_{i-1}}^{x_i}
{\rm ln}\frac{x_0y_0}{x_0^2+y_0^2}dxdx',
\end{eqnarray*}
where $x_0 = L-x+x'\cos\theta$, and $y_0 = -x'\sin\theta$. The 
equilibrium structure of the dislocation can be obtained by minimizing
the total energy with respect to the dislocation density.  

To contrast and understand different cross-slip behavior in Al and Ag, we 
have carried out {\it ab initio} calculations for the $\gamma$-surface of 
Ag while the $\gamma$-surface of Al has been published elsewhere \cite{Lu1}.
A supercell containing six layers in the [111] direction is used to calculate
the $\gamma$-surface for Ag. The {\it ab initio} calculations are based 
on the pseudopotential plane-wave method
\cite{Payne} with a kinetic energy cutoff of 55 Ry for the plane-wave
basis and a $k$-point grid consisting of (16,16,4) divisions along the
reciprocal lattice vectors. Owing to the planar nature of dislocation
core structure of fcc metals, we disregard the displacement perpendicular 
to the
slip planes and partially consider the shear-tension coupling by
performing volume relaxation along the [111] direction in the $\gamma$-surface
calculations. We present the complete $\gamma$-surface
for Ag in Fig. 2. The most striking difference between the
$\gamma$-surface of Ag and Al is the vast difference in intrinsic stacking
fault energy, which is 165 mJ/m$^2$ for Al and 14 mJ/m$^2$ for Ag. 
This dramatic difference in $\gamma$-surface gives rise to very different 
dislocation core structures
and cross-slip behavior that we are going to explore.

The model calculation is set up by introducing a screw dislocation at the 
intersection of the two slip planes without applying external stress 
to the system at first. The initial
configuration of the dislocation is specified by a step function for the screw 
displacement $f^{\rm I}_3(x)=0$ for $x<L$ and $f^{\rm I}_3(x)=b$ for $x \geq L$. 
All other displacement components including those on the cross-slip
plane are set to zero initially. This corresponds to a
pure screw dislocation with a zero width ``spread'' on the primary plane.
We then relax the dislocation structure according to the energy 
functional. The Burgers vector of Ag, $b$ = 2.84 \AA~ is determined from 
{\it ab initio} calculations 
and elastic constants are chosen from experimental values \cite{Hirth}. 
The corresponding parameters of Al have been given elsewhere \cite{Lu1}. 
Having determined all the parameters entering the model, we obtain the equilibrium 
structure of the dislocations, represented by their density 
$\rho(x)$ shown in Fig. 3. 
The screw dislocation in Al which starts out at the primary plane spontaneously 
spreads into the cross-slip plane, as the density peak at the cross-slip
plane indicates. As expected, the edge component of the density is zero 
at the cross-slip plane because only screw displacement can cross-slip.
On the other hand, the screw dislocation in Ag
dissociates into two partials, separated by 7.8 $b$ ($\approx$ 22 \AA)
distance. These partial dislocations cannot cross-slip, as the arrows 
indicate, without first annihilating their edge components, and 
the dislocation density on the cross-slip plane is essentially zero. The
partial separation distance we obtained from the model calculation
is in excellent 
agreement with the TEM measurement for that 
in Ag, which is about 20 \AA~ \cite{Cockayne}. Apparently,
the lack of obvious dissociation in Al stems from the fact that Al
has a much higher intrinsic stacking fault energy than that of Ag. 

In order to examine the stress effect on dislocation core structure 
and cross-slip process, we apply external Escaig stress to
the dislocation. The Escaig stress with pure edge component 
interacts only with 
the edge component of dislocation densities, 
extending or shrinking the stacking fault width depending on its sign. 
The results of the
partial separation as a function of Escaig stress are summarized in
Fig. 4. Without external stress, the partial separation is 7.8 $b$ 
for Ag and zero for Al. Under positive Escaig stress, 
the partial separation rises rapidly for Ag whereas it remains 
zero in Al until the stress reaches the threshold to separate the 
overlapping partials. To activate 
cross-slip, however, one needs to apply a negative Escaig stress to
annihilate the edge components of the partials' displacement, known 
as constriction process. Upon application of negative stress, the partials 
in Ag move towards each other and reduce the width of stacking
fault. During this process, the edge components of displacement 
from the two partials annihilate each other while the screw components 
are being built up. 
However there is a lower limit 
for the separation that one can achieve, which is 1.7 $b$ for Ag. 
This is in agreement with the atomistic simulations for Cu, 
reporting a corresponding value of 1.6 $b$ \cite{Duesbery}. In the wake of
the (partial) constriction process, a pure screw dislocation segment
is formed at the intersection of the two planes, which can cross-slip.
On the other hand, further increasing the negative stress does not 
complete the constriction but 
rather increases the partial separation. This is due to the fact that the
remaining edge components of the partials interact with the stress, and
as a result the two partials exchange signs and move away from
each other until the lattice breaks down. 
   
We have also estimated critical energetics that 
are relevant to cross-slip. For example, we calculated
constriction energy defined as the difference in dislocation
core energy between the normal and constricted states. By approximating
the state with 1.7 $b$ separation between partials as the constricted state,
 we were able to estimate the
constriction energy for Ag to be 0.05 eV/$b$. Obviously the constriction 
energy for Al is zero because its normal state is fully constricted. 
We have also
calculated the critical stress for cross-slip which is defined as the
glide stress in the cross-slip plane to move a partially constricted 
dislocation from the primary plane to the cross-slip plane 
\cite{Duesbery}. The critical stress for 
cross-slip in Ag is found to be 0.0105 eV/\AA$^3$, comparing to 0.0020
eV/\AA$^3$ in Al. Finally we estimated cross-slip energy 
barrier which in the context of our calculations is defined as the
difference in dislocation core energy before and after cross-slip
takes place by applying the above mentioned critical stress for
cross-slip. In other words, we calculate the core energy difference
for the dislocation between its normal state and the state that the dislocation
just starts to cross-slip under the critical cross-slip stress. 
Under this definition, we find the cross-slip energy barrier for Ag as
0.31 eV/$b$, much greater than that of Al, which is 0.05 eV/$b$. Our
result for the cross-slip energy barrier should not be compared directly 
to the corresponding experimental
value because the dislocations are assumed to be straight in our current 
implementation of the Peierls-Nabarro model. However it is possible to
extend the present formalism to deal with an arbitrarily curved dislocation
where a more realistic cross-slip energy barrier can be obtained. 
Nevertheless the present model is still sufficient to provide reliable energetics 
for straight dislocations. 
 
To conclude, we have presented a novel model that can treat dislocation
cross-slip and constriction based on the Peierls-Nabarro framework. The
$\gamma$-surface entering the model is determined from {\it ab initio}
calculations which provide reliable atomic interactions across the
slip plane. Using this model, we find that the screw dislocation in
Al can spontaneously spread into cross-slip plane while the same dislocation
in Ag splits into partials and cannot cross-slip. The dislocation
response to external stresses is studied, and in particular negative Escaig 
stresses are applied to the dislocation to simulate constriction process.
We find that it is impossible to achieve 100\% constriction for 
straight partial dislocations.
By computing dislocation core energy in different stress states, we
are able to estimate dislocation constriction energy for Al and Ag. We have 
also calculated critical stress and energy barrier for dislocation cross-slip,
and from which we confirm that dislocation 
cross-slip is much easier in Al than in Ag. Since our {\it ab initio} model 
calculation is much faster than direct {\it ab initio} atomistic simulations, 
the model will 
be invaluable for alloy design where the 
goal is to select ``right'' elements with ``right'' composition 
for an alloy to have 
desired dislocation properties, such as cross-slip properties.

\begin{acknowledgments}
Two of us (G.L. and N.K.) acknowledge the support from  
Grant No. DAAD19-00-1-0049 through the U.S. Army Research Office. G.L. 
was also supported by Grant No. F49620-99-1-0272 through the U.S. Air 
Force Office for 
Scientific Research. We thank E. Kaxiras for suggesting this 
interesting problem to us.
\end{acknowledgments}

\begin{figure}
\includegraphics[width=300pt]{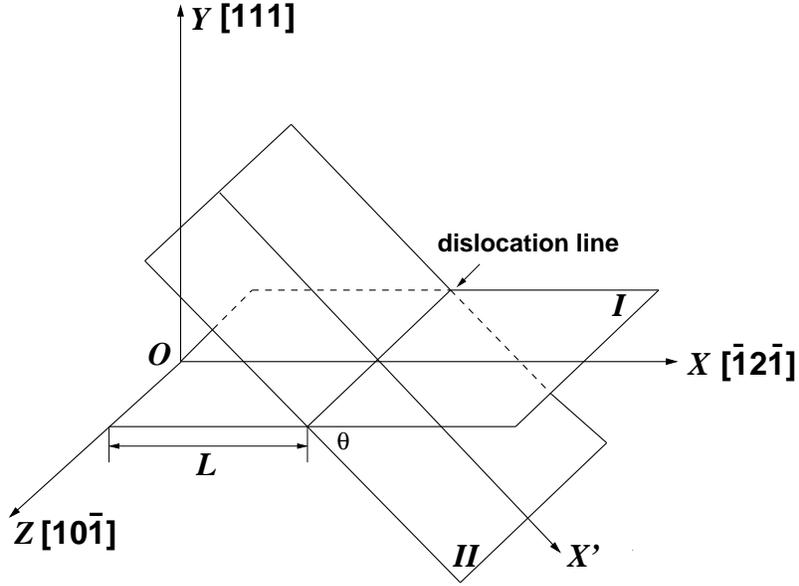}
\caption{Cartesian set of coordinates showing the directions relevant 
to the screw dislocation
located at the intersection of the two slip planes.}
\label{fig1}
\end{figure}

\begin{figure}
\includegraphics[width=300pt]{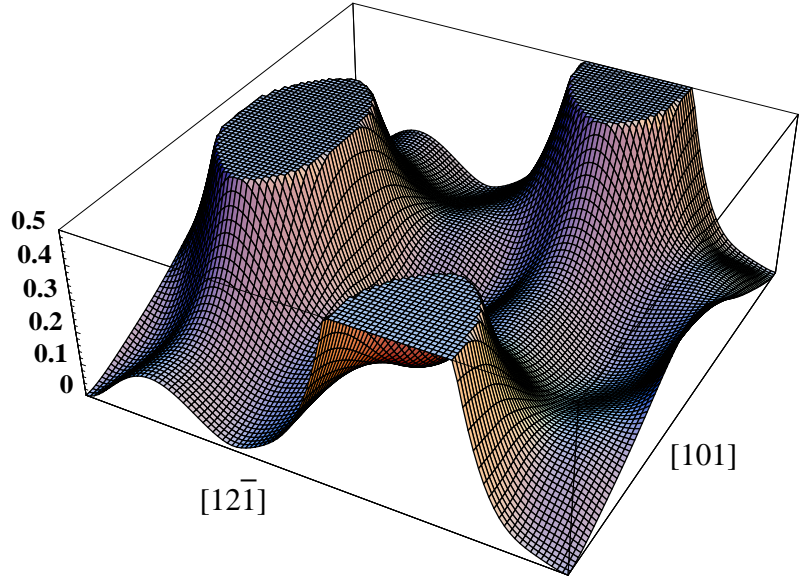}
\caption{The $\gamma$-surface (J/m$^2$) for displacements along a (111) plane for Ag.
The corners of the plane and its center correspond to identical equilibrium
configuration, i.e., the ideal lattice. The $\gamma$-surface is truncated  
to emphasize the more interesting region.}
\end{figure}

\begin{figure}
\includegraphics[width=300pt]{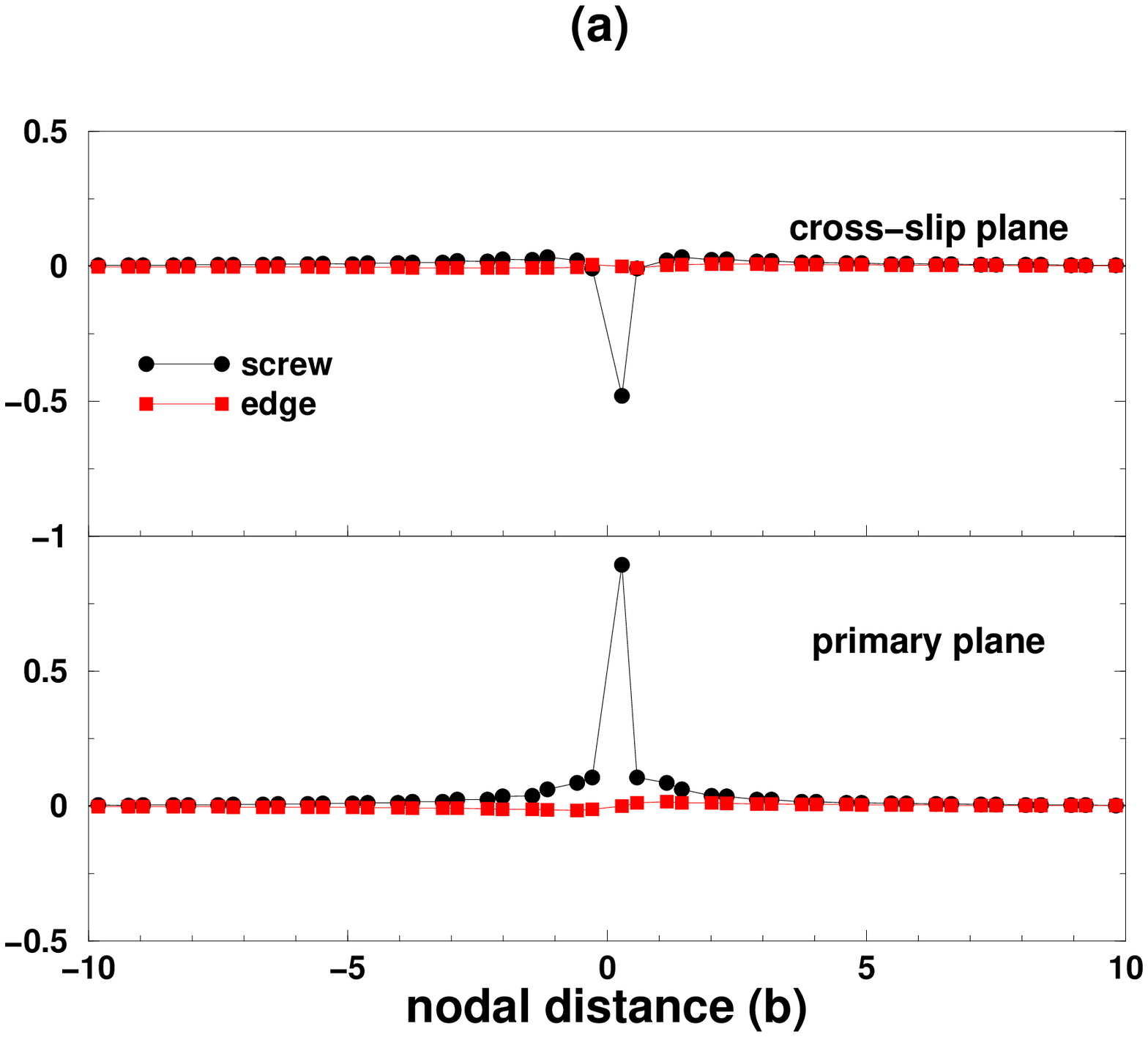}
\includegraphics[width=300pt]{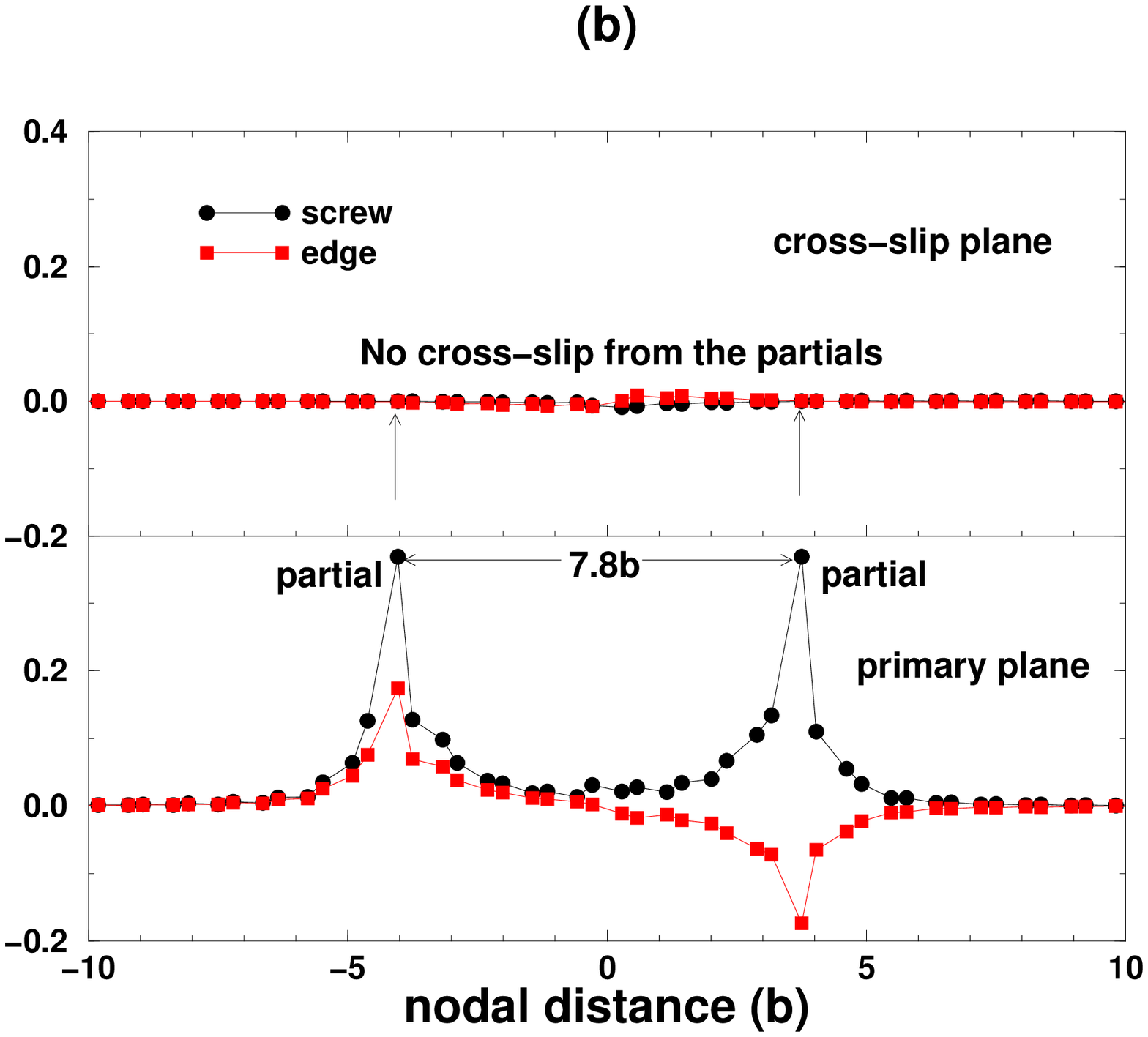}
\caption{Dislocation displacement density $\rho$(x) for Al (Fig.3(a)) and Ag (Fig. 3(b)). 
The peaks in the density plot represent (partials) dislocations.}
\end{figure}

\begin{figure}
\includegraphics[width=300pt]{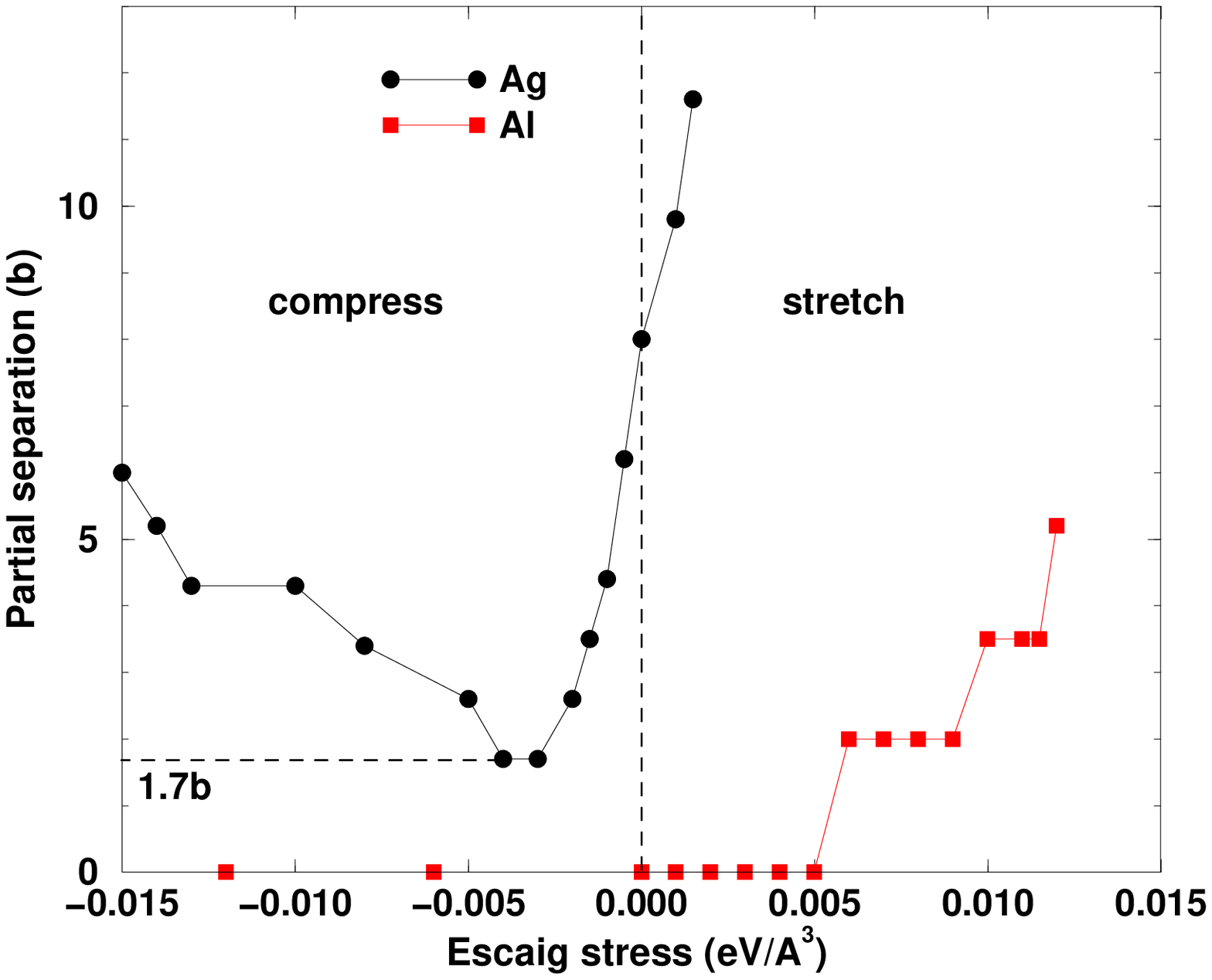}
\caption{Partial separation as a function of applied Escaig stress. The vertical dashed line
represents the zero stress separating the compress and stretch regions. The horizontal dashed
line indicates the minimal separation distance for Ag.} 
\end{figure}


\begin{thebibliography}{99}
\bibitem{Friedel}
J. Friedel, in {\it Dislocations and Mechanical properties of Crystals},
edited by J.C. Fisher (Wiley, New York, 1957).
\bibitem{Escaig}
B. Escaig, in {\it Dislocation Dynamics}, edited by A.R. Rosenfeld {\it et al.}
(McGraw-Hill, New York, 1968).
\bibitem{Rasmussen} 
T. Rasmussen, K.W. Jacobsen, T. Leffers, O.B. Pederson, S.G. Srinivasan,
and H. J$\acute{o}$nsson, Phys. Rev. Lett. {\bf79}, 3676 (1997).
\bibitem{Rao}
S. Rao, T.A. Parthasarathy, and C. Woodward, Philos. Mag. A {\bf79}, 1167 
(1999).
\bibitem{Joos}
B. Jo$\acute{o}$s, Q. Ren and M.S. Duesbery, Phys. Rev. B {\bf50}, 5890 (1994).
\bibitem{Juan}
Y.M. Juan and E. Kaxiras, Philos. Mag. A {\bf74}, 1367 (1996).
\bibitem{Bulatov}
V. V. Bulatov and E. Kaxiras, Phys. Rev. Lett. {\bf78}, 4221 (1997).
\bibitem{Hartford}
J. Hartford, B. von Sydow, G. Wahnstr$\ddot{o}$m, and B.I. Lundqvist, Phys.
Rev. B {\bf58}, 2487 (1998).
\bibitem{Lu1}
G. Lu, N. Kioussis, V. V. Bulatov, and E. Kaxiras, Phys. Rev. B {\bf62}, 3099 (2000);
Philos. Mag. Lett. {\bf80}, 675 (2000).
\bibitem{Lu3}
G. Lu, Q. Zhang, N. Kioussis and E. Kaxiras, Phys. Rev. Lett. {\bf87}, 095501 (2001). 
\bibitem{Nabarro}
F.R.N. Nabarro, Adv. Phys., {\bf 1}, 269 (1952).
\bibitem{Payne}
M.C. Payne, M.P. Teter, D.C. Allan, T.A. Arias and J.D. Joannopoulos, Rev. Mod. Phys.,
{\bf 64}, 1045 (1992).
\bibitem{Hirth}
J.P. Hirth and J. Lothe, {\it Theory of Dislocations}, 2nd ed. (Wiley, New York, 1992).
\bibitem{Cockayne}
D.J.H. Cockayne, M.L.Jenkins and I.L.F. Ray, Philos. Mag. {\bf 24}, 1383 (1971).
\bibitem{Duesbery}
M.S. Duesbery, Modelling Simul. Mater. Sci. Eng. {\bf 6}, 35 (1998).
\end{thebibliography}
\end{document}